\documentclass{PoS}

\title{Newtonian and relativistic polytropes}

\ShortTitle{Newtonian and relativistic polytropes}

\author{\speaker{F. M. Araujo}\thanks{To Universidade Federal do ABC and CNPq for the financial support and possibility of development of a research project.}\\
        Centro de Matemática, Computação e Cognição (CMCC), Universidade Federal do ABC (UFABC), Santo André - SP, Brazil\\
        E-mail: \email{fefis.araujo@uol.com.br}}

\author{C. B. M. H. Chirenti\\
        Centro de Matemática, Computação e Cognição (CMCC), Universidade Federal do ABC (UFABC), Santo André - SP, Brazil\\
        Max-Planck-Institut für Gravitationsphysik, Albert Einstein Institut, Golm, Germany\\
        E-mail: \email{cecilia.chirenti@ufabc.edu.br}}

\abstract{The equation of state inside very compact objects like neutron stars is still largely unkown. Even though a lot progress has been made in recent years to develop the so-called realistic equations of state, a lot of insight can be gained by using polytropic equations of state to integrate the stellar equations of structure. In this work we provide a brief review of the Newtonian and relativistic equations of structure and present some numerical results, which we believe that can be useful for students starting to work on this field. The internal structure of the Newtonian polytropes is obtained by the numerical integration of the Lane-Emden equation, and we used our results which can also be used to study the Chandrasekhar limit for white dwarfs. However, Newtonian physics cannot correctly describe very compact and massive objects. There is an upper mass limit for neutron stars, supported by observations, which is not predicted by the Newtonian equations. Neutron stars are best described under the framework of General Relativity. The introduction of the TOV equations (as well as the relativistic Lane-Emden equations) is, therefore, necessary to correctly identify stable and non-stable models via the mass-radius relation. The analysis of the EoS also becomes relevant to discard models that allow a possible violation of causality (sound speed larger than $c$).

}

\FullConference{4th School on Cosmic Rays and Astrophysics,\\
		August 25- September 04, 2010\\
		Sao Paulo Brazil}

\begin{document}

\section{Introduction}
\label{sec:introduction}

The current period of discoveries in the fields of Physics allows the acquisition of knowledge and understanding of physical phenomena previously unknown. Cosmology, for instance, is no different. The accelerated expansion of the universe, the recurrent related terms energy and dark matter, the possible emission of gravitational waves in binary systems among others are the new phenomenology present nowadays. Amongst this phenomenological diversity, one can also highlight an object of fundamental importance to the physical verification of gravitational waves emission: neutron stars. Thus, the study of these compact objects is important for the substantiation of predictions made based on the theory of General Relativity \cite{Schutz}. 

 Accurate models of neutron stars are still under development. The equation of state of matter in the extreme conditions found in the center of those objects remains unknown. There is a variety of factors that can be taken into account. The existence of a solid crust, magnetic fields, rotation and superfluidity are just a few of them\cite{Schutz}.
The study of neutron stars is therefore highly interdisciplinary, covering many areas of Physics. Nevertheless, it is necessary to check whether models derived from different assumptions and simplifications are physically consistent and correct. Due to its high compactness it is natural that neutron stars are studied under the framework of General Relativity.

In this work we have studied the non-relativistic (Newtonian) and relativistic stellar equations of structure with a polytropic equation of state, and numerical results were obtained with a MATLAB code. Although these results are not new in the literature, we believe that the compilation of the relevant equations and some numerical results will be useful for students starting on this field.

The paper is organized as follows. Section~\ref{sec:theory} briefly review the relevant equations for stellar structure used in this paper. In Section~\ref{sec:results} we present the numerical results obtained for relativistic and non-relativistic polytropes and finally in Section~\ref{sec:conclusions} we present our final remarks.

\section{Stellar equations of structure}
\label{sec:theory}

\subsection{Non-relativistic stars}

A non-relativistic star in equilibrium can be described by the equations
\begin{eqnarray}
\label{SE1}
\frac{dP}{dr} &=& -\frac{GM(r)\rho(r)}{r^2}\,,\\
M(r) &=& \int_0^r 4\pi \rho r^2 dr\,,
\label{SE2}
\end{eqnarray}
where $r$ is the radial coordinate inside the star, $P$ is the density, $\rho$ is the total mass density and $M(r)$ is the total mass inside a radius $r$. An equation of state $P(\rho)$ also needs to be supplied, and we use a polytropic equation of state of the form
\begin{equation}
P = K\rho^{\gamma}\,, \quad \gamma = 1+\frac{1}{n}
\label{EOS}
\end{equation}
where $n$ the polytropic index and $K$ may be established by specifying the density and pressure at the center.

Introducing the new variables $x$ and $y$ defined by the system
\begin{eqnarray}
\label{x}
r &=& \frac{x}{A}\,,\\
\rho&=& \rho_cy^n(x) \,,
\label{y}
\end{eqnarray}
where 
\begin{equation}
A = \left[\frac{4\pi G\rho_c^\frac{n-1}{n}}{(n+1)K}\right]^{1/2}\,,
\label{A}
\end{equation}
one can use eqs.~(\ref{SE1}), (\ref{SE2}) and (\ref{EOS}) to obtain the Lane-Emden equation:
\begin{equation}
\frac{1}{x^2}\frac{d}{dx}\left( x^2\frac{dy}{dx} \right) + y^n = 0\,.
\label{LE}
\end{equation}
The Lane-Emden functions are the solutions of eq.~(\ref{LE}) with initial conditions $y(0) = 1$ and $\frac{dy}{dx}(0) = 0$. On the surface of the star $x(R) = AR$ and $y(R) = 0$. 

After solving the Lane-Emden equation (\ref{LE}) for a given value of the polytropic index $n$, we can fix the radius $R$ and the total mass $M(R)$ of the star. The parameter $A$ is fixed by the relation $x(R) = AR$, the central density is given by
\begin{equation}
\rho_c = -\frac{M(R) A^3}{4\pi x^2(R)y'(R)}\,,
\label{rho_c}
\end{equation} 
and the constant $K$ can be obtained from eq.~(\ref{A}). Finally, $\rho(r)$ and $p(r)$ are given by eqs.(\ref{y}) and (\ref{EOS}).

\subsection{Relativistic stars}

In the relativistic treatment, eq.~(\ref{SE1}) is replaced by the Tolman-Oppenheimer-Volkoff (TOV) equation given by
\begin{equation}
-\frac{dP}{dr} = \frac{GM + 4\pi Gr^3P/c^2}{r^2(1-2GM/c^2r)}\left(\rho + \frac{P}{c^2}\right)\,.
\label{TOV}
\end{equation}

The relativistic Lane-Emden equation can be obtained by introducing the new dimensionless coordinates (similarly to what was done in the non-relativistic case):
\begin{eqnarray}
\label{Rxi}
\xi &=& Ar\,,\\
\label{Rv}
v(\xi) &=& \frac{A^3}{4\pi\rho_c}m(r)\,,\\
\label{Rtheta}
\theta^n(\xi) &=& \frac{\rho}{\rho_c}\,.
\end{eqnarray}

With the definitions (\ref{Rxi}), (\ref{Rv}) and (\ref{Rtheta}) and the equation of state (\ref{EOS}), eqs.~(\ref{SE2}) and (\ref{TOV}) can be rewritten as
\begin{eqnarray}
\label{RLE1}
\frac{d\nu}{d\xi} &=& \xi^2\theta^n\,,\\
-\xi^2\frac{d\theta}{d\xi} &=& \frac{(\nu + \sigma\theta\xi d\nu / d\xi)(1 +\sigma\theta)}{1 - 2\sigma (n+1)\nu/\xi}\,,
\label{RLE2}
\end{eqnarray}
where 
\begin{equation}
\sigma = \frac{P_c}{\rho_c c^2} = \frac{K\rho^{1/n}}{c^2}\,,
\label{sigma}
\end{equation}
and $P_c$ is the central pressure. In the non-relativistic limit $\sigma \to 0$, eqs.~(\ref{RLE1}) and (\ref{RLE2}) can be recombined to give the Lane-Emden equation~(\ref{LE}) with $x = \xi$ and $y = \theta$.

The relativistic Lane-Emden functions are the solutions of eqs.~(\ref{RLE1}) and (\ref{RLE2}) with the initial conditions $\theta(0) = 1$ and $v(0) = 0$. The surface of the star is represented by the first zero $\xi_1$ of $\theta$.

Having solved eqs.~(\ref{RLE1}) and (\ref{RLE2}) for given values of $n$ and $\sigma$, one can choose a value of $\rho_c$ to fix the other parameters of the solution and the equation of state.

\section{Numerical results}
\label{sec:results}

We developed a MATLAB code to solve the Lane-Emden equation~(\ref{LE}) and obtain the stellar structure inside the polytrope \cite{Maciel}. Some relevant quantities obtained for the Lane-Emden functions with different values of $n$ are listed on table~\ref{LE-Table}.

\begin{table}[t]
\centering
\begin{tabular}[htp!]{c|c|c}
$n$ & $x(R)$ & $y'(R)$ \\
\hline 
0 & 2.4502 &  -0.8167\\
1 & 3.1400 & -0.1061\\
2 & 4.3526 & -0.1273\\
3 & 6.8966 & -0.0424\\
4 & 14.9716 & -0.008\\
\hline
\end{tabular}
\caption{Values obtained for $x(R)$ and $y'(R)$ from the integration of the Lane-Emden equation.}
\label{LE-Table}
\end{table}

\begin{figure}[!hbtp]
\centering
\includegraphics[width=0.45\textwidth]{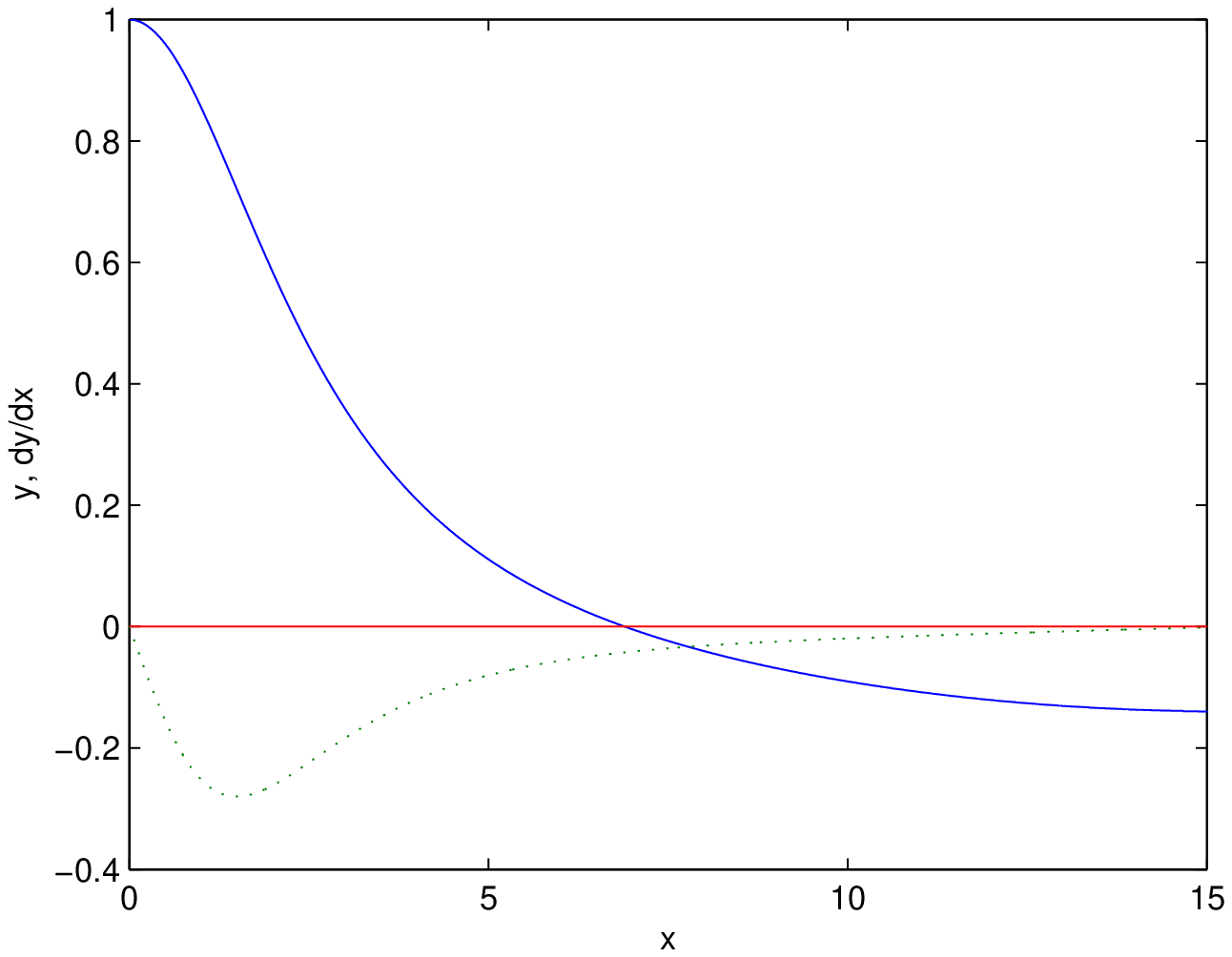}
\includegraphics[width=0.45\textwidth]{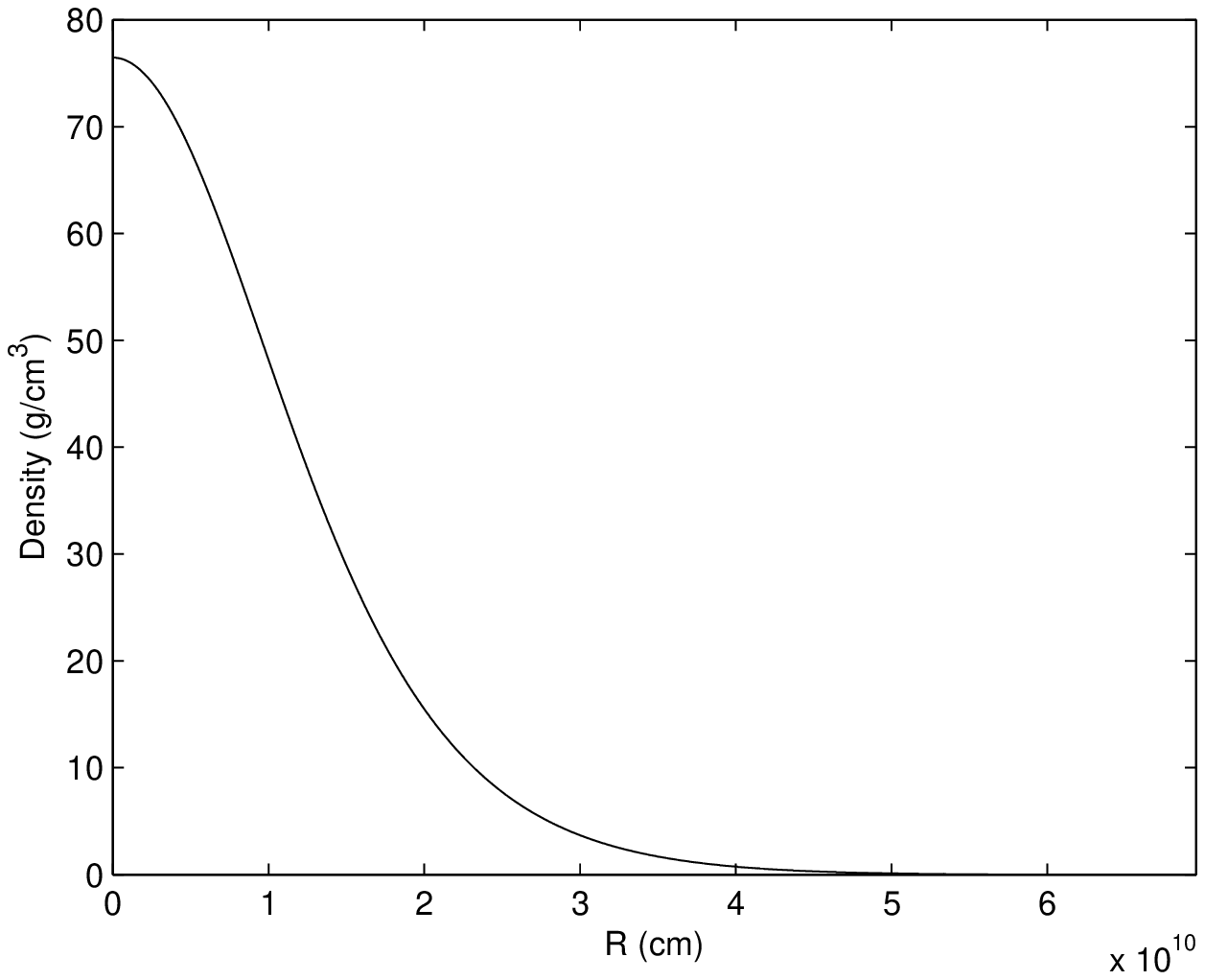}
\includegraphics[width=0.45\textwidth]{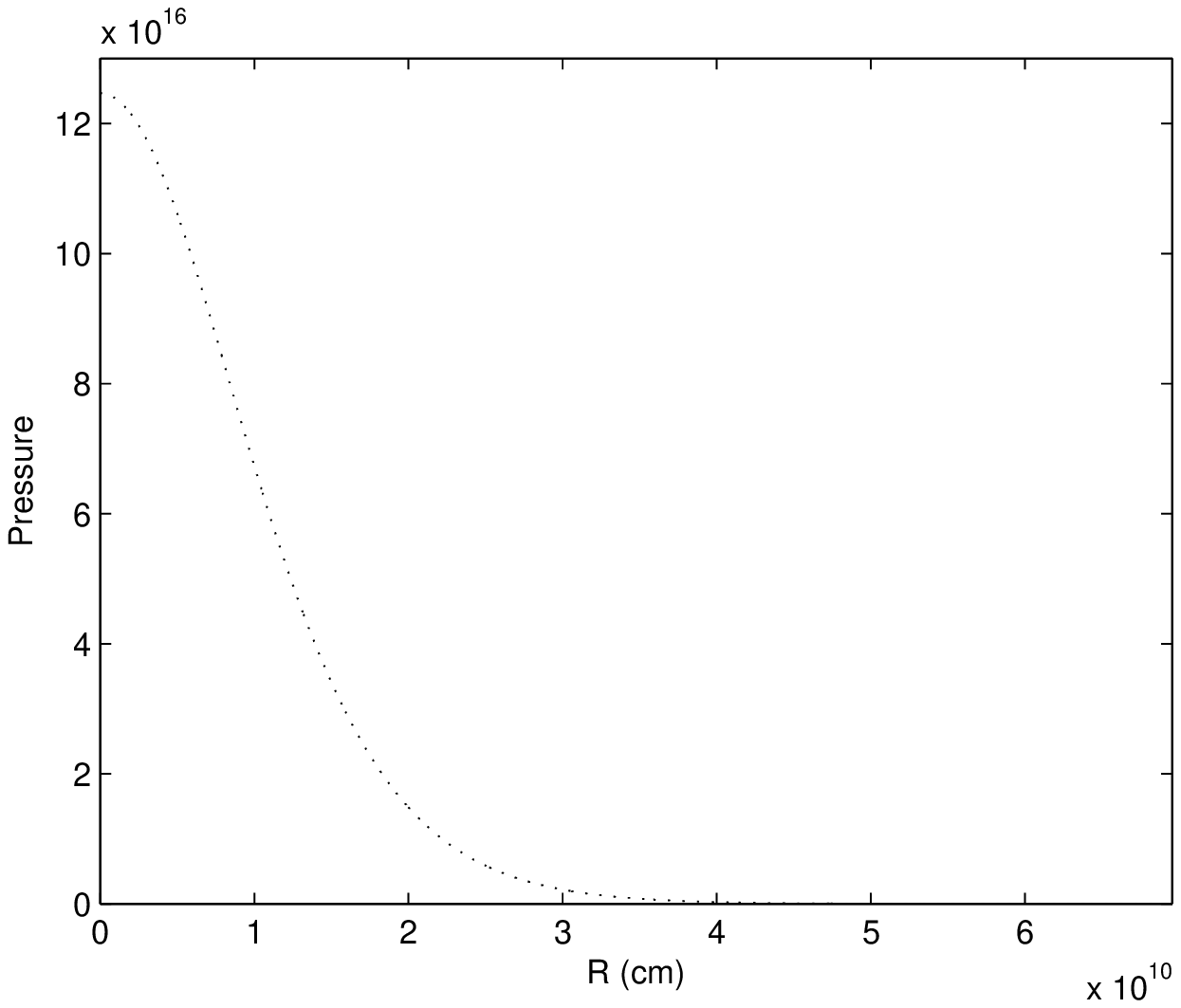}
\includegraphics[width=0.45\textwidth]{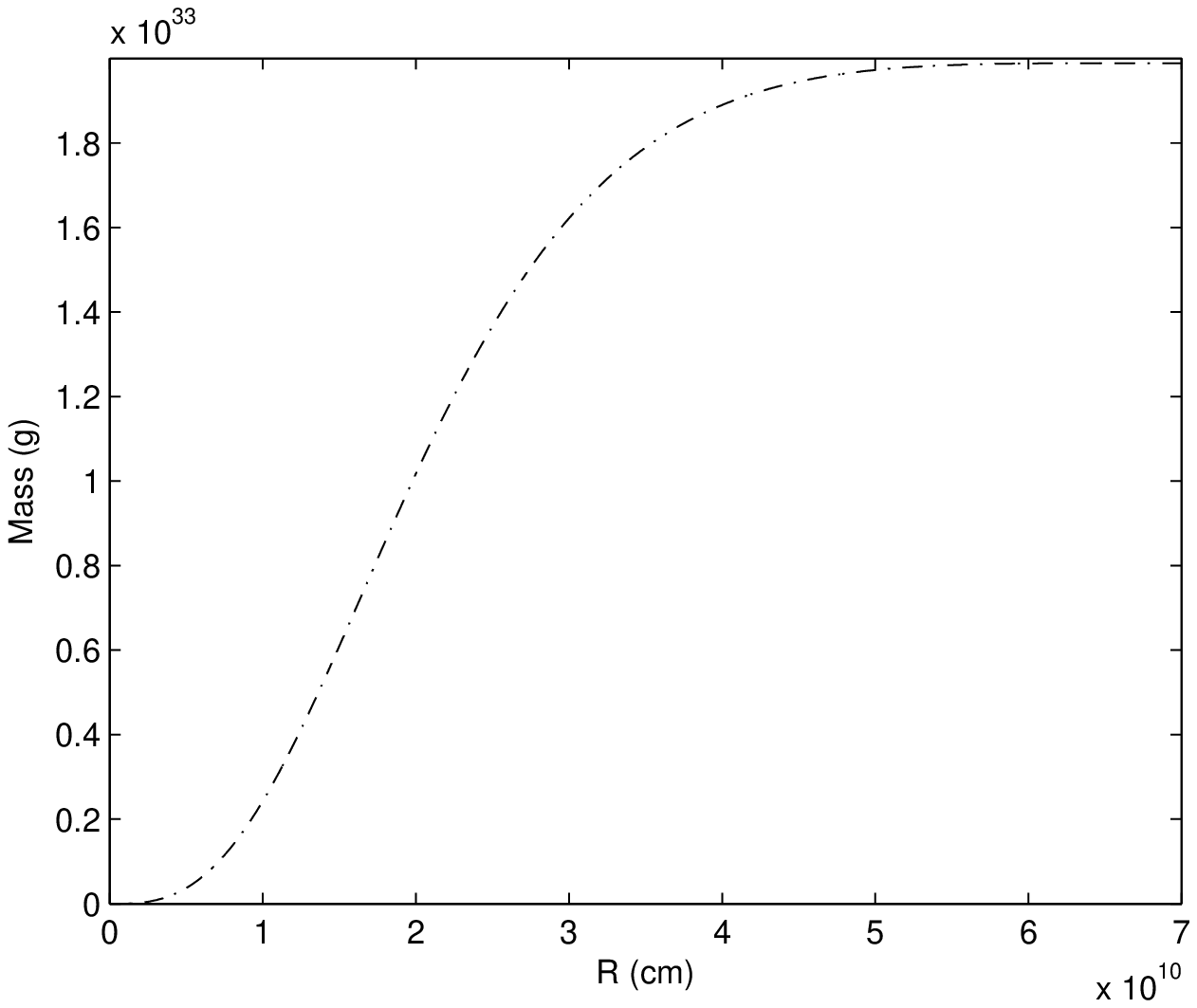}
\caption{Numerical results obtained for $y(x)$ (solid line) and $dy/dx$ (dotted line) with $n = 3$, and radial behavior of the density $\rho$, pressure $p$ and mass $M$ of a star with $1 M_{\odot}$ and $1 R_{\odot}$.}
\label{FIG1}
\end{figure}

On figure~\ref{FIG1}, we present some typical results obtained for a $n=3$ polytrope (usually refered to as the standard model), representing a star of $1 M_{\odot}$  and $1 R_{\odot}$, with $\rho_c = 76,46\, g/cm^3$ and $K = 3.84 \times 10^{14}\, dyn\, cm^2/g^{4/3}$. We point out that a white dwarf can also be modelled as a $n = 3$ polytrope. In a white dwarf, the electron degeneracy pressure provides the means for stabilizing the star against its gravitational radiation, and the equation of state of a relativistic degenerate electron gas is of the from $p = K\rho^{4/3}$. Note that from eqs.~(\ref{rho_c}) and (\ref{A}) it can be shown that for $n = 3$ the total mass $M(R)$ is independent from the central density $\rho_c$, which corresponds to the Chandrasekhar limit for white dwarfs \cite{Ryden}. These stars have a mass less than the Chandrasekhar limit of approximately $1.4 M_{\odot}$ and radius of the order of $10^4$ km \cite{Silbar}.

After a low or medium mass star has finished burning most of its hydrogen and helium, it can become a white dwarf. At this stage, if the star has a mass below that of the Chandrasekhar limit, the collapse is limited by the degeneracy pressure of electrons (the Pauli exclusion principle), resulting in a stable white dwarf. However, if the star does not have any more fuel to maintain its production of energy and has a mass greater than the Chandrasekhar limit, the pressure exerted by the electrons is not sufficient to withstand the force of gravity and the star collapses. Its density will increase dramatically, leading to the formation of a neutron star or black hole \footnote{It is also possible the formation of a quark star. However, this is a theoretical solution and will not be addressed in this project.} \cite{Narlikar,Inverno}. The Chandrasekhar limit is a result from the effects predicted by Quantum Mechanics considering the behavior of electrons that cause the degeneracy pressure of a white dwarf. Electrons, as fermions, cannot occupy the same energy level, i.e., cannot be described by the same quantum numbers. Thus, in white dwarfs, a large amount of electrons are at higher energies, causing a certain amount of pressure capable of sustaining the star. For a mass greater than the limit, the degenerescence pressure becomes insufficient to prevent the imminent contraction of the star. So far, no white dwarf with mass greater than the Chandrasekhar limit has been observed \cite{Chung}.

In the relativistic treatment, we have extended our calculations, including all the results previously done, to stiffer (incompressible matter, $n \to 0$) EoS. The results obtained are presented in tables~\ref{table1}-\ref{table3}. For any solution with given values of $n$, $\sigma$ and and a chosen value of $\rho_c$, we can determine $K$ from eq.~(\ref{sigma}), and we can obtain $R$ and $M(R)$ with

\begin{eqnarray}
R &=& A^{-1}\xi_1\,,\\
\tilde{M} &\equiv& \sigma^{\frac{3-n}{2}}\nu (\xi_1)\,,\\
M &=& \frac{4\pi\rho_c}{A^3}\nu (\xi_1) = \left[\frac{(n+1)c^2}{4\pi G}\left(\frac{K}{c^2}\right)^n\right]^{1/2}\tilde{M}\,.
\end{eqnarray}

From these results it is possible to verify which solutions are stable and causal. The velocity of low frequency sound waves is given by $v_s^2 = \frac{\mathrm{d}P}{\mathrm{d}\rho} = \frac{\gamma P}{\rho} = c^2\gamma (P/\rho c^2)$. At the stellar center where $P/\rho c^2$ achieves its maximum value $\sigma$, $v_s^2 = c^2\gamma\sigma$. If $\sigma > 1/\gamma$, the speed of sound would exceed that of light and the fluid inside the star could violate causality. 

\begin{figure}[!hbtp]
\centering
\includegraphics[width=0.6\textwidth]{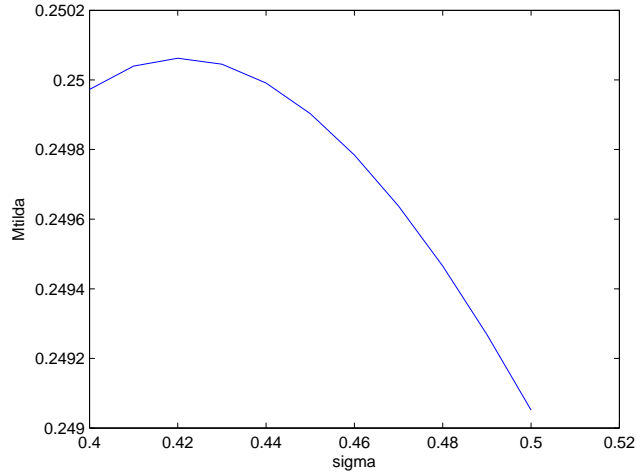}
\caption{$\tilde{M}$ as a function of $\sigma$ for $n = 1$. We can see that $\sigma_{CR}\approx 0.42$ separates stable ($\sigma < \sigma_{CR}$) and unstable models ($\sigma > \sigma_{CR}$)}.
\label{FIG2}
\end{figure}

Both $\tilde{M}$ and $M$ increase with $\sigma$ until a value $\sigma_{CR}$, as can be seen in figure~\ref{FIG2} for $n = 1$. Since $dm/d\rho_c > 0$ is a necessary condition for stability, $\sigma_{CR}$ marks the onset of the first mode of radial instability \cite{Bludman}.

\begin{table}[t]
\centering
\begin{tabular}[htp!]{c|c|c|c||c|c|c|c|}
$\sigma$ & $\xi_1$ & $\nu(\xi_1)$ & $\tilde{M}$ & $\sigma$ & $\xi_1$ & $\nu(\xi_1)$ & $\tilde{M}$\\
\hline 
0.0 & 6.8966   & 2.0182 & 2.0182 & 0.0 & 4.3531 & 2.411  & 0.0   \\
0.1 & 6.8264   & 1.0785 & 1.0785 & 0.1 & 3.6989 & 1.2987 & 0.4107\\
0.2 & 7.9494   & 0.713 & 0.713   & 0.2 & 3.3983 & 0.8403 & 0.3758\\
0.3 & 10.8346  & 0.5386 & 0.5386 & 0.3 & 3.271  & 0.6055 & 0.3316\\
0.4 & 17.8244  & 0.4516 & 0.4516 & 0.4 & 3.2473 & 0.468  & 0.296 \\
0.5 & 37.2163  & 0.4214 & 0.4214 & 0.5 & 3.2967 & 0.38 & 0.2687  \\
0.6 & 91.0674  & 0.4493 & 0.4493 & 0.6 & 3.3986 & 0.3201 & 0.2479\\
0.7 & 162.6175 & 0.5266 & 0.5266 & 0.7 & 3.5469 & 0.2773 & 0.232 \\
0.8 & 187.233  & 0.5969 & 0.5969 & 0.8 & 3.7334 & 0.2457 & 0.2197\\
0.9 & 187.029  & 0.6375 & 0.6375 & 0.9 & 3.9539 & 0.2216 & 0.2102\\
1.0 & 183.6571 & 0.657 & 0.657   & 1.0 & 4.2027 & 0.203  & 0.203 \\
\end{tabular}
\caption{Lane-Emden relativistic functions parameters for $n=3.0$ (left) and $n = 2$ (right).}
\label{table1}
\end{table}

\begin{table}[t]
\centering
\begin{tabular}[htp!]{c|c|c|c||c|c|c|c|}
$\sigma$ & $\xi_1$ & $\nu(\xi_1)$ & $\tilde{M}$ & $\sigma$ & $\xi_1$ & $\nu(\xi_1)$ & $\tilde{M}$\\
\hline 
0.0 & 3.1416 & 3.1416 & 0.0    & 0.0 & 2.7523 & 3.786 & 0.0    \\
0.1 & 2.5989 & 1.7514 & 0.1751 & 0.1 & 2.2898 & 2.1735 & 0.1222\\
0.2 & 2.277  & 1.1426 & 0.2285 & 0.2 & 2.0008 & 1.4358 & 0.192 \\
0.3 & 2.0642 & 0.8192 & 0.2457 & 0.3 & 1.8013 & 1.035 & 0.2298 \\
0.4 & 1.9133 & 0.6249 & 0.25   & 0.4 & 1.6544 & 0.791 & 0.2516 \\
0.5 & 1.8008 & 0.4981 & 0.2491 & 0.5 & 1.5409 & 0.6295 & 0.2647\\
0.6 & 1.7143 & 0.4101 & 0.246  & 0.6 & 1.4503 & 0.5164 & 0.2727\\
0.7 & 1.6458 & 0.3461 & 0.2423 & 0.7 & 1.3759 & 0.434 & 0.2779 \\
0.8 & 1.5903 & 0.2979 & 0.2383 & 0.8 & 1.3136 & 0.3717 & 0.2812\\
0.9 & 1.5447 & 0.2605 & 0.2344 & 0.9 & 1.2604 & 0.3231 & 0.2833\\
1.0 & 1.5066 & 0.2307 & 0.2307 & --- & ---    & ---    & ---   \\
\end{tabular}                  
\caption{Lane-Emden relativistic functions parameters for $n=1.0$ (left) and $n = 0.5$ (right).}
\label{table2}
\end{table}

\begin{table}[t]
\centering
\begin{tabular}[htp!]{c|c|c|c||c|c|c|c|}
$\sigma$ & $\xi_1$ & $\nu(\xi_1)$ & $\tilde{M}$ & $\sigma$ & $\xi_1$ & $\nu(\xi_1)$ & $\tilde{M}$\\
\hline
0.0 & 2.504  & 4.6089 & 0.0     & 0.0 & 2.4508 & 4.9071 & 0.0   \\
0.1 & 2.1037 & 2.7308 & 0.09689 & 0.1 & 2.0642 & 2.9318 & 0.0927\\
0.2 & 1.8438 & 1.8377 & 0.1781  & 0.2 & 1.8113 & 1.9807 & 0.1772\\
0.3 & 1.6593 & 1.3367 & 0.2333  & 0.3 & 1.6306 & 1.4452 & 0.2375\\
0.4 & 1.5202 & 1.0254 & 0.2716  & 0.4 & 1.4938 & 1.111  & 0.2811\\
0.5 & 1.4108 & 0.8172 & 0.2991  & 0.5 & 1.3856 & 0.8867 & 0.3135\\
0.6 & 1.322  & 0.671  & 0.3199  & 0.6 & 1.2976 & 0.7283 & 0.3385\\
0.7 & 1.2482 & 0.5634 & 0.3359  & 0.7 & 1.2241 & 0.6114 & 0.3581\\
0.8 & 1.1856 & 0.4816 & 0.3485  & 0.8 & 1.1616 & 0.5225 & 0.3739\\
0.9 & 1.1317 & 0.4176 & 0.3585  & 0.9 & 1.1078 & 0.4532 & 0.3869\\
--- & ---    & ---    & ---     & 1.0 & 1.0606 & 0.3977 & 0.3977
\end{tabular} 
\caption{Lane-Emden relativistic functions parameters for $n=0.1$ (left) and $n = 0.0$ (right).}
\label{table3}
\end{table}

\section{Final remarks}
\label{sec:conclusions}

The present work enabled the acquisition of fundamental knowledge to develop models of Newtonian and Relativistic stars. Through the implementation of numerical methods along with literature review, it was possible to obtain results that simulate the intra-stellar medium and to verify the main physical properties of polytropic stars. As for the simulations, the results obtained show a very good agreement when compared to previous results found in literature.
The classic study of the polytrope of index $n = 3$ allowed us to determine important physical characteristics such as the maximum mass limit and relate it to a phenomenon well known to the white-dwarfs, the Chandrasekhar limit. From this, it was  possible to understand the concept of degenerate matter and its relation to gravitational collapse.

However, the verification that Newtonian equations do not correctly explain the behavior of stellar masses for different radii has become essential. Newtonian equations describe the behavior of masses through an indefinite growth whereas General Relativity provides a maximum value of mass for each equation of state. It is, therefore, necessary to use relativistic corrections for an accurate simulation of compact and massive objects.

The execution of the simulations allowed, therefore, the development of technical and intuitive notions about the general behavior of stars, understanding the importance of the introduction of relativistic corrections and the results arising therefrom, which are essential for determining subluminal and stable solutions.


\begin{thebibliography}{99}
\bibitem{Schutz} B. Schutz, Gravity from the ground up, Cambridge, 2003.
\bibitem{Press} W.H. Press, B.P. Flannery and S.A. Teukolsky, Numerical Recipes in Fortran 77: The Art of Scientific Computing, Cambridge, 1986. 
\bibitem{Maciel} W.J. Maciel, Introdução à estrutura e evolução estelar, Edusp, São Paulo, 1999.
\bibitem{Silbar} R.R. Silbar and S. Reddy, Neutron stars for undergraduates students, Am. J. Phys. 72, 7 (2004).
\bibitem{Ryden} B. Ryden, Introduction to Cosmology, Pearson Addison Wesley, San Francisco, 2003
\bibitem{Narlikar} J.V. Narlikar, An Introduction to Cosmology, Cambridge, 2002.
\bibitem{Inverno} R. D'Inverno, Introducing Einstein's relativity, New York: Oxford, 1992.
\bibitem{Chung} K. C. Chung, Introdução à Física Nuclear, Rio de Janeiro: Eduerj, 2001.  
\bibitem{Bludman} S.A. Bludman, Stability of General-Relativistic Polytropes, Am. J. Phys. 183, 7 (1973).
\end{thebibliography}
\end{document}